\documentclass{article}

\PassOptionsToPackage{numbers}{natbib}

\usepackage[preprint]{neurips_data_2024}

\usepackage[utf8]{inputenc} 
\usepackage[T1]{fontenc}    
\usepackage[hidelinks]{hyperref}       
\usepackage{url}            
\usepackage{booktabs}       
\usepackage{amsfonts}       
\usepackage{nicefrac}       
\usepackage{microtype}      
\usepackage{xcolor}         

\usepackage{tikz}         
\usepackage{comment}
\usepackage{dirtree}

\usepackage{subfigure}
\usepackage{paralist}
\usepackage[scale=0.94]{sourcecodepro}
\usepackage[T1]{fontenc}

\title{\texttt{posteriordb}: Testing, Benchmarking and Developing Bayesian Inference Algorithms}

\author{%
  Måns Magnusson\thanks{Corresponding author: \url{mans.magnusson@statistik.uu.se}} \\
  Dept.\ of Statistics\\
  Uppsala University\\
  Uppsala, Sweden
  \And
  Jakob Torgander\\
  Dept.\ of Statistics\\
  Uppsala University\\  
  Uppsala, Sweden \\  
  \And
  Paul-Christian Bürkner \\
  Dept.\ of Statistics \\
  TU Dortmund University \\
  Dortmund, Germany
  \And 
  Lu Zhang\\
  Dept.\ of Population and \\ Public Health Science\\
  University of Southern California\\
  Los Angeles, CA, USA\\
  \And   
  Bob Carpenter  \\
  Center for \\
  Computational Mathematics \\
  Flatiron Institute\\  
  New York, NY, USA \\      
  \And   
  Aki Vehtari  \\
  Dept. of Computer Science\\
  Aalto University\\  
  Helsinki, Finland \\    
}

\begin{document}

\maketitle

\begin{abstract}
The generality and robustness of inference algorithms is critical to the success of widely used probabilistic programming languages such as Stan, PyMC, Pyro, and Turing.jl.  When designing a new general-purpose inference algorithm, whether it involves Monte Carlo sampling or variational approximation, the fundamental problem arises in evaluating its accuracy and efficiency across a range of representative
target models. To solve this problem, we propose \texttt{posteriordb}\footnote{\url{https://github.com/stan-dev/posteriordb/tree/v1.0}}
, a database of models and data sets defining target densities along with reference Monte Carlo draws. We further provide a guide to the
best practices in using \texttt{posteriordb} for model evaluation and comparison. To provide a wide range of realistic target densities, \texttt{posteriordb} currently comprises 120 representative models and has been instrumental in developing several general inference algorithms. 
\end{abstract}

\section{Introduction}

The \texttt{posteriordb} package was developed to address the general problem of evaluating probabilistic inference algorithms of the variety found in a probabilistic programming language (PPL).  In this section, we review why the problem and our approach to solving it with \texttt{posteriordb}.

\paragraph{Probablistic Programming Languages}  Probabilistic programming languages (PPLs) are (often embedded) domain-specific languages programming languages for probabilistic modelling.  PPLs have attracted hundreds of thousands of users over the past three decades. These frameworks enable users to flexibly specify models involving data and unknown parameters and provide general probabilistic inference conditioned on data (e.g., parameter estimation, event probability estimation, predictive inference).  Based on academic paper citations, PPLs are used in almost every corner of applied statistics and machine learning, including the physical, biological and social sciences, medicine, engineering, education, finance, and entertainment. The most widely used PPLs today \cite[according to ][]{strumbelj2024past} are Stan \cite{carpenter2017stan}, Tensorflow Probability \cite{dillon2017tensorflow}, PyMC \cite{salvatier2016probabilistic}, Pyro \cite{bingham2019pyro}, JAGS \cite{plummer2003jags}, and Turing.jl \cite{tarek2020dynamicppl}.

PPLs support both the development and programming of statistical models and also provide general inference to support downstream applications. For the PPLs listed above, this is carried out in a ``black box'' fashion that relies only on the model's log density and gradients, not on the specific structure of the model.  Given a (not necessarily normalized) joint probability model $p(y,\theta)$ for the unknown parameters $\theta$ and observed data $y$, the main interest is typically to estimate expectations based on $\theta$ such as parameter estimates, event probabilities, or predictions.  Starting out, we specify an unnormalized joint model, $p(\theta,y)$ using PPL syntax. Given data $y$, Bayes's theorem says the posterior of interest is proportional to the joint density, which can be unpacked into the likelihood times the prior, $p(\theta \mid y) \propto p(\theta, y) = p(y \mid \theta) \cdot p(\theta)$.  Using the posterior density, we can calculate the posterior predictive distribution $p(\tilde{y} \mid y)$ for new data $\tilde{y}$, estimate event probabilities $\Pr[\theta \in E],$ and estimate parameters as $\mathbb{E}[\theta \mid y]$ \cite{gelman2013bayesian}.

In most settings, computing $p(\theta|y)$ is analytically intractable, so we lack direct access to the posterior. 
PPLs instead use approximate inference algorithms.

Recently, interest has been focused on "black-box" inference algorithms applicable across diverse models. Several such inference algorithms have been proposed, such as Hamiltonian Monte Carlo \cite[HMC, ][]{neal2011mcmc,hoffman2014no}[], variational inference \cite[VI, ][]{jordan1999introduction,ranganath2014black,blei2017variational}, and Laplace approximations \cite[LA]{tierney1986accurate,rue2009approximate} along with various adaptations and improvements of these approaches \cite{bales2019selecting,dhaka2021challenges,wu2022foundation,modi2024variational,wang2024stein}. These black-box inference algorithms have different properties. HMC, as other Markov chain Monte Carlo (MCMC) methods, will, in many settings, converge to the posterior in total variation distance \cite{tierney1994markov}. MCMC algorithms are generally computationally costly. On the other hand, VI and LA can be less computationally costly but often introduce a bias in estimating the posterior expectations \cite{wang2018frequentist}, introducing a trade-off between accuracy and computational cost.

\paragraph{\texttt{posteriordb}} Given the different inference algorithms and PPL implementations, algorithm adaptations, initialization and general improvements, it is challenging to determine how these proposed methods compare. Commonly, new methods are evaluated on a small number of posteriors, making it hard to know how well specific inference algorithms work in general. When developing and maintaining PPLs and inference algorithms, ideally, we would like to \emph{test} that they work as intended for many posteriors. When developing new algorithms, we want to \emph{assess performance} to gain insights on which posteriors the algorithms work well and where they fail. Finally, we want to \emph{benchmark} suggested algorithms to assess how newly proposed methods compared to already existing approaches. 

We introduce \texttt{posteriordb}, a database to aid in algorithms and PPL development, such as testing, performance assessment and benchmarking. To facilitate this, \texttt{posteriordb} contains a collection of hundreds of posteriors, models, datasets, and reference posteriors in a simple repository structure (see Figure \ref{fig:posteriordb} and \ref{fig:posteriordb_structure}). The database also includes references to papers, details about the posteriors, and metadata on models and data to simplify performance analysis.  \texttt{posteriordb} is a fully open posterior database, and we encourage contributors to share their posteriors and models with the repository, especially more complicated posteriors. Today, \texttt{posteriordb} has already been used in multiple studies on posterior approximations \cite[e.g., see ][]{dhaka2021challenges,welandawe2022robust, baudart2021compiling,liang2022fat}.

\begin{figure}
\centering
    \begin{minipage}{0.45\textwidth}
        \centering
    
 	\tikzstyle{block} = [rectangle, draw, fill=black!20, 
 	text width=6em, text centered, rounded corners, minimum height=2em]
 	\tikzstyle{line} = [draw, -stealth]
 	
 	\begin{tikzpicture}[node distance = 2cm, auto]\label{ams1}
 	\node [block] (posterior) {Posterior (1)};
  	\node [block, above left of=posterior] (model) {Model (2)};
  	\node [block, above right of=posterior] (data) {Data (3)};
 	\node [block, below of=posterior] (reference) {Reference Draws (4)};

 \path[line] (posterior) -- (model);
 \path [line] (posterior) -- (data);
 \path [line] (posterior) -- (reference);
 	\end{tikzpicture}
\caption{\it The conceptual \texttt{posteriordb}}
\label{fig:posteriordb}
    \end{minipage}\hfill
    \begin{minipage}{0.45\textwidth}
\footnotesize
   \dirtree{%
.1 posterior\_database.
.2 posteriors.
.2 models.
.3 stan.
.3 pymc3.
.3 info.
.2 data.
.3 data-raw.
.3 data.
.3 info.
.2 reference\_posteriors.
.3 draws.
.4 draws.
.4 info.
.3 summary\_statistics.
.4 mean\_value.
.5 mean\_value.
.5 info.
.4 mean\_value\_squared.
.5 \texttt{...}.
.2 bibliography.
}
        \caption{{\it Structure of} \texttt{posteriordb}}
        \label{fig:posteriordb_structure}
    \end{minipage}
\end{figure}

Until now, previous work on collecting models and datasets has been focused on particular sub-classes of models. Examples of such collections are causal structure graph models \cite{rios2021benchpress} and Bayesian neural networks \cite{vadera2022ursabench}. In addition, most of the popular PPLs currently provide example models for comparison and evaluation purposes, such as Stan \cite{stan2021ref}, BUGS and JAGS \cite{lunn2000winbugs, plummer2003jags}, PyMC(3) \cite{salvatier2016probabilistic}, (Num)Pyro \cite{bingham2019pyro, phan2019composable}, Turing.jl \cite{tarek2020dynamicppl}, ADMB/TMB \cite{monnahan2018no}, and NIMBLE \cite{de2017programming}, to name a few, not to mention black-box MCMC packages that do not rely on PPLs or gradients, like emcee \cite{foreman2013emcee}. Some small examples of sets of posteriors for more general benchmarking are \texttt{Inference Gym} \cite{inferencegym2020} and \texttt{PPLbench} \cite{pplbench2022,kulkarni2020ppl}. Many of these posteriors are also included in \texttt{posteriordb}.

Section 2 introduces the main use cases of posterior repositories. Section 3 introduces \texttt{posteriordb} and how it can be used. Section 4 describes an example of using \texttt{posteriordb} in evaluating the Pathfinder algorithm \cite{zhang2022pathfinder} and Section 5 concludes.

\section{Use cases of databases of Bayesian posteriors}
\label{sec:use_cases}

The main interest in most Bayesian inference is posterior expectations. Let $\hat{p}(\theta \mid  y)$ be any approximation of the posterior $p(\theta \mid y)$.  For example, with variational inference, $\hat{p}$ will be a member of the variational family, with Laplace approximation, $\hat{p}$ will be multivariate normal, and with sampling, $\hat{p}$ will be the discrete empirical distribution.
Further, we assume that it is possible to generate a set $\{\theta_{(s)} \}_{s=1}^S$ of $S$ draws $\theta^{(s)} \sim \hat{p}(\theta \mid y)$ for $s \in \{ 1, \ldots, S \}$ with which to compute expectations and quantiles for empirical evaluation.  While many different expectations might be of interest, the focus is commonly on means, variances, and tail quantiles of parameters and predictive variables defined as transforms of parameters and data (e.g., posterior predictions and event probability forecasts).

Assessing the performance of an inference algorithm is non-trivial. Although, from a general perspective, we can evaluate inference algorithms in three ways, all of which may be measured using \texttt{posteriordb} targets.
\begin{description}
    \item[Accuracy] How well does the algorithm approximate the target density (e.g., KL-divergence, squared error, Wasserstein distance, etc.)? 
    \item[Efficiency] What is the computational cost (in time, flops, gradient evaluations, memory, power consumption, etc.), of the algorithm?
    \item[Generality] Which classes of posteriors and posterior inference problems can the algorithm solve with what accuracy and efficiency?
\end{description}

\subsection{Testing algorithms and their implementations}

Testing posterior inference algorithms poses more problems than standard software testing \cite{dutta2018testing} and shares
similarities with functional testing \cite{kaner1999testing}. 
When testing posterior inference algorithms, especially asymptotically unbiased algorithms, such as MCMC and HMC, the focus is usually on testing the posterior expectations. Let
$
\epsilon^2_{\hat{p},g} = \left(\strut\mathbb{E}_p(g(\theta) \mid  y)-\mathbb{E}_{\hat{p}}(g(\theta) \mid  y)\right)^2
$
be the squared approximation error for a given expectation.
Then, the marginal means and variances of the posterior have the benefit that, if they are finite, the Markov chain central limit theorem can be used to assess the inference algorithm \cite{jones2004markov}. If the algorithm works as expected, the approximation error $\epsilon_{\hat{p},g}$ will decrease over iterations at the rate $\mathcal{O}(1 / \sqrt{n})$. Hence, we can use a high-quality reference posterior approximation for testing purposes; see Section~\ref{sec:reference_posterior_definition} for details.

A \emph{testably correct algorithm} generates draws whose marginal distribution follows the target density and thus can be used to evaluate inference algorithms.  Independent samplers are testably correct, as are MCMC methods run for finite amounts of time given some verifiable assumptions, such as geometric ergodicity \cite{roberts1997geometric}. 
However, with challenging posterior density geometry, computational limitations can result in low accuracy for finite runs of MCMC.  For example, random-walk Metropolis, Gibbs, and HMC all fail to sample the funnel density \cite{neal2003slice} in finite time, despite asymptotic guarantees, because of the poor and varying condition in the mouth and neck \cite{papaspiliopoulos2007general,modi2023delayed}. Nevertheless, we have two ways out of this dilemma.  First, we can reparameterize, which allows us to take independent draws for the funnel example.  Second, we can assess a poorly mixing or even asymptotically biased algorithm in terms of how well it estimates expectations in finite time.

In terms of evaluation reliability, the best we can do is analytically known expectations, as we can derive in many cases (for example, the funnel mean parameter values are all zero).  The next best thing to do is to take independent samples, the standard error for which is known.  The last resort is to take MCMC draws and attempt to verify the results are correct (e.g., with simulation-based calibration \cite{talts2018validating,modrak2023simulation}) and then thin them until roughly independent.

Even though posterior expectations are the main statistic of interest, various divergences between distributions can be used for a more holistic assessment of the general properties of the posterior, such as Wasserstein distance \cite{villani2009optimal,craig2016exponential}, maximum mean discrepancy \cite{JMLR:v13:gretton12a}, or the Pareto-$\hat{k}$ diagnostic \cite{Vehtari+etal:2024:PSIS}. To assess accuracy, we propose the following:
\begin{compactenum}
    \item RMSE of posterior moments of interest compared to a reference posterior true analytic moments or high accuracy moment estimates based on a trusted algorithm and high computational budget,  
    \item Wasserstein divergence between distribution approximation and true posterior,
    \item Maximum Mean Discrepancy (MMD) between the approximation and the true posterior, and
    \item Pareto-$\hat{k}$ diagnostic for the density ratio indicating whether importance sampling can be used to correct $\hat{p}(\theta \mid y)$ to approximate $p(\theta \mid y)$.
\end{compactenum}

Some inference algorithms, such as variational inference or Laplace approximations, are biased in most applications (meaning non-zero expected error).  They can also blow up the problem's dimensionality by directly modeling covariance.  Hence, accuracy becomes more important to assess how well the true posterior is approximated for these algorithms. Again, we can judge the accuracy using posterior expectations or more holistic approaches.

When testing a posterior inference algorithm for correctness and accuracy, a large set of posteriors that are easy to run simplifies the task.  With posteriors of different shapes, sizes, and geometries, and hence difficulty, \texttt{posteriordb} allows developers to get a handle on an algorithm's performance in a wide range of realistic settings.

When testing posterior parameter estimates, evaluating the estimates of parameters squared is good practice.  Accurate estimates of squared parameters are useful because it's the component required for estimating variance through $\textrm{var}[\theta] = \mathbb{E}[\theta^2] - \mathbb{E}^2[\theta].$  With algorithms like HMC, it is possible to do extremely well at estimating parameter expectations while poorly estimating parameters squared.

\subsection{Development of new algorithms}

The second use case of a repository of posterior distributions is the development of new posterior approximation algorithms. When developing new inference methods, some algorithms may work for certain posteriors and fail for others. For example, HMC has difficulties with funnels, and normal approximation methods work best for approximately multivariate normal posteriors. We want to find out for which type of posteriors a new algorithm works well and when it fails. Hence, many different posteriors can be used both to find unknown failure cases and demonstrate expected difficulties.

When developing posterior approximation algorithms, one of the more important aspects is the accuracy-computation trade-off. 
Assessing the computational performance of posterior approximation algorithms can be implementation \emph{independent} implementation or \emph{dependent}.  An implementation-dependent quantity is wall time or energy consumed; implementation-independent quantities are floating-point operations per second (flops), log density evaluations, or gradient evaluations.  Typically, computation is dominated by log density and/or gradient calculations; with automatic differentiation, the log density and gradient are computed simultaneously \cite{margossian2019review}.

We can compare the accuracy after a fixed amount of computation when developing algorithms, whether they are biased or asymptotically exact.  Here, examples of implementation-dependent measures are log density evaluation per second (LDE/s), gradient evaluations per second (GE/s), as well as the effective sample size per second (ESS/s). For algorithms producing draws, ESS/s, where effective sample size can be estimated from standard deviation estimated over several runs and standard error as $\textrm{ESS} = (\textrm{sd} / \textrm{se})^2$.  This is the central implementation-dependent performance metric since it measures the approximation precision achievable within a given practical time budget \cite{burkner_models_2023}.

\subsection{Benchmarking of existing algorithms}

The computation-accuracy trade-off shows the importance of comparing and benchmarking algorithms. As part of methods development, we want to make informed decisions on which algorithms to implement and use, that is, we want to benchmark methods. This also applies to minor but important improvements of existing algorithms, such as warmup adaptations and computational improvements.

Development and benchmarking require challenging models and posteriors for which we don't necessarily have efficient algorithms yet, or the current algorithms cannot reach the asymptotic regime in a feasible time. Benchmarking a large number of posteriors is crucial to appropriately evaluate the breadth of posteriors that can be approximated with high accuracy and assess the associated computational costs. Even if an asymptotically unbiased algorithm and implementation are correct and work well for certain posterior geometries, they might fail spectacularly for others.
For example, dynamic HMC with fixed step size integrator fails to reach the asymptotic regime for many funnel-shaped posteriors in feasible time \cite{betancourt2015hamiltonian}. On the other hand, Laplace and variational approximations combined with importance sampling can reach the asymptotic regime for some of these same funnel-shaped posteriors if they are sufficiently low dimensional \cite{yao2018yes}.

A large set of posterior distributions, such as funnel-shaped posteriors, multimodal posteriors, discrete and discrete-continuous-mixed posteriors, high-dimensional posteriors, finite and infinite posteriors (Dirichlet Processes), large data posteriors, and simple, analytically tractable posteriors, can assess the generality of algorithms. A posterior approximation algorithm can be useful if it works well for some models and can be diagnosed when it doesn't work. The generality of algorithms also describes what type of posteriors individual algorithms have performance issues. Hence, we want to know the types of errors and problems to assess the generality of benchmarked algorithms.

\subsection{Development and maintenance}
The process of testing algorithms includes multiple steps. Employing the same rigorous approaches used for previously well-tested algorithms is common practice when developing a new algorithm. This ensures that the new implementation meets the expected standards of functionality and reliability. Similarly, when maintaining existing software, testing serves the dual purpose of verifying that changes haven't compromised the integrity of the inference algorithm and that the algorithm's performance remains unaffected. Regression testing, as it's known in computer science, compares algorithm outputs over the development lifecycle to catch any behavior or performance ``regressions.''

\section{\texttt{posteriordb}: a database for testing, benchmarking, and development}
\label{sec:posteriordb}

We introduce \texttt{posteriordb} with all the above-mentioned use cases in mind. It is a comprehensive repository containing posteriors, models, data, and reference posteriors. The primary objective is to leverage this set of posteriors to test, assess, benchmark, develop and maintain PPLs and posterior approximation algorithms. 
The database contains both more difficult/complex posteriors, such as Covid-19 epidemic models \cite{flaxman2020report}, Bayesian neural networks \cite{lampinen2001bayesian}, and simpler, standard posteriors, such as the eight schools example \cite{rubin1981estimation,gelman2013bayesian}. All posteriors, data and models are stored in the same format, simplifying the estimation of many posteriors for generality and benchmarking purposes. 

\subsection{The \texttt{posteriordb} components}
\label{sec:objects}
The \texttt{posteriordb} contains four main types of objects (see Figure \ref{fig:posteriordb} and \ref{fig:posteriordb_structure} for an overview). 

The \emph{posterior} (1) object summarizes all information about a specific posterior in the collection. A posterior object points to a (not necessarily normalized) joint model $p(y,\theta)$, data $y$, and a reference posterior (if any). The purpose of separating models from data is that some models use the same data, which is relevant for model comparison diagnostics and benchmarking, and some models can be used for multiple datasets, enabling model comparison diagnostics and benchmarking. Finally, the posterior points to a reference posterior if such a reference posterior exists.

The \emph{model} (2) object in \texttt{posteriordb} stores an (unnormalized) joint model, $p(y,\theta)$, in the form of PPL code and JSON information files. While most models are currently written in Stan, the structure allows us to easily include code from other PPLs, such as PyMC, Tensorflow Probability, Pyro, etc.

The \emph{data} (3) objects, $y$, are stored as compressed JSON files to simplify and facilitate the use of the data. Each data file also contain an information JSON file. The \texttt{data-raw} folder contains code and information on processing the data in case processing has been done.

The \emph{reference posterior} (4) object (in the JSON sense of ``object'') represents the true posterior distribution, usually in the form of posterior draws, if it is possible to compute such a representation. To serve as a reference posterior, a set of draws must be of very high quality, as detailed further in Section \ref{sec:reference_posterior_definition}. Depending on the size and the possibility of computing the posterior distribution, the reference posterior draws themselves and/or the corresponding posterior expectations are stored in the reference posterior object as compressed JSON files. The benefit of a true, or well approximated, reference posterior is that we can assess, to a given tolerance and a specific computational budget, if the output of an algorithm is correct concerning a true underlying posterior distribution. An information JSON file includes exact details of how the posterior draws were computed.

As an example of a posterior in \texttt{posteriordb}, we can look at the \texttt{eight\_\allowbreak{}schools-\allowbreak{}eight\_\allowbreak{}schools\_\allowbreak{}centered} posterior that points to the \texttt{eight\_schools} dataset and the centered parametrization of the eight school model \texttt{eight\_schools\_centered} \cite{betancourt2015hamiltonian}. Further, the posterior object includes the posterior dimension and points to the reference posterior \texttt{eight\_\allowbreak{}schools-\allowbreak{}eight\_\allowbreak{}schools\_\allowbreak{}noncentered}. The centred parametrization is well-known to have difficulties due to the funnel geometry of the posterior. Hence, the non-centred parametrization was used to compute a reference posterior for the centred model.

The choice of including models follows the principle of the goal of including a large and diverse set of posteriors. We also focus on including posteriors where data and models have been published openly. This enables users of \texttt{posteriordb} to read up more on specific models and data in more detail, since references are included in the repository.

Accessing \texttt{posteriordb} can be done in two ways. The first 
way is to directly access the content of \texttt{posteriordb} from the repository \url{https://github.com/stan-dev/posteriordb}. The folder \texttt{posteriord\_database} in the repository contains the data in a folder structure that can be seen in Figure \ref{fig:posteriordb_structure}. Data and reference posterior draws are compressed as zip archives. It is also possible to interact with the \texttt{posteriordb} through the R package (\url{https://github.com/stan-dev/posteriordb-r}) and the Python library \url{https://github.com/stan-dev/posteriordb-python}) to simplify quick access. All posteriors, data, models, reference posteriors, and software are version-controlled using semantic versioning.

\subsection{Additional metadata}
We include potential relevant meta-information in all four objects, the number of parameters for posterior objects, and keywords to simplify the selection of posteriors. We also include keywords for posteriors to enhance the capacity to assess diverse performance aspects, enabling a comprehensive understanding of algorithmic behaviours and for benchmarking purposes, for example, developing new algorithms. This also aids in diagnosing issues with new algorithms or in benchmarking settings. Further, if available, posteriors, models, and data also contain bibliographical entries to 
to provide further knowledge about the posteriors. 

\subsection{The reference posterior object}
\label{sec:reference_posterior_definition}
A key component of \texttt{posteriordb} is the \textit{reference posterior} (RP) object. The RP object consists of (approximately) independent and identically distributed Monte Carlo draws from the corresponding posterior \textit{model} object and serves as a representation of the \textit{true} underlying posterior distribution. Depending on the form of the posterior $p(\theta \mid y )$, the draws of the RP object are created in one of two ways, by independent sampling or through MCMC.

For some simple models, it is possible to compute the true posterior analytically and then sample from it, but this is not the case for most posteriors. We expand the set of reference posteriors by including models for which we have high confidence that we can obtain draws of sufficiently high quality that pass as coming from the true posterior. We use MCMC, specifically Stan's implementation of NUTS, to obtain reference posteriors for well-behaved models from which we cannot independently sample.  First, we compute a set of draws, $\{\theta^{(s)}\}$ and include them in \texttt{posteriordb}. Second, we compute the posterior parameter expectations (either analytically or with the draws). The posterior means support direct error evaluation, and the draws allow more holistic evaluation, for example, using Wasserstein distances. The inclusion of draws further aids in identifying areas and specific types of posteriors that exhibit suboptimal performance and regions of difficulty.  

Even with the 10\,000 roughly independent draws supplied by \texttt{posteriordb}, the standard error for a parameter estimate will be the parameter's standard deviation divided by 100 (square root of 10\,000). For instance, estimating the mean of a standard normal distribution (which is 0) will have a standard error of 0.01. This imposes an upper bound on the accuracy of a system being evaluated before the error in the reference dominates the estimated error.

We define a reference posterior, or expectations thereof, as 10~000 draws or more from the true posterior distribution. In practice, this is only possible in a limited number of analytical settings. In the case of MCMC, the chains should be thinned so that the draws are approximately independent to make further comparisons easier. However, we also use MCMC to generate draws from the true posterior distribution. To consider draws generated using MCMC as a reference posterior, we require \begin{compactenum}
    \item 10 000 draws per parameter in the model (or more),
    \item approximately independent draws, that is, all parameters have a mean autocorrelation at lag 1 over the chains that is less than 0.05 in absolute value,  
    \item an $\hat{R}$
    below 1.01 for all parameters, \cite[see ][]{vehtari2021rank}    
    \item all expected fraction of missing information (E-FMI) is below 0.2 \cite[see ][]{betancourt2017conceptual}, and
    \item if HMC is used, no divergent transitions \cite[see ][]{betancourt2016diagnosing}.
\end{compactenum}
To compute the reference posterior draws using MCMC, we use Stan HMC/NUTS. However, other approaches, such as model-specific algorithms, can also be used in special circumstances if it is clear that it is necessary, for example, for discrete parameter models.

Our repository also includes interesting and challenging posteriors, even if these posteriors lack a reference posterior when this is not possible or impractical to compute. This is, for example, the case in combinatorially multimodal posteriors such as high-dimensional clustering models like latent Dirichlet allocation models \cite{blei2003latent} or Bayesian neural networks.

\subsection{The current scope of \texttt{posteriordb}}

\texttt{posteriordb} currently contains 147 posteriors, 120 models, 91 datasets and 46 reference posterior draws. Of these, roughly a third are easy cases, whereas the remaining two-thirds are more challenging. Table~\ref{tab_complex_posteriors} contains examples of posteriors that can be run using Stan HMC but where standard use results in a large number of divergent transitions or leapfrog steps, indicating more complex posterior geometries. Note that higher HMC acceptance rates can improve the divergence rates, and the max tree depth bounds the mean number of leapfrog steps. Hence, the results can rather be seen as an indication of sampling difficulties of some of the posteriors in \texttt{posteriordb}. 

We can see that some posteriors, such as \texttt{soil\_carbon-soil\_incubation}, produce divergent transitions, indicating large changes in curvature. In contrast, \texttt{synthetic\_grid\_RBF\_kernels-kronecker\_gp} needs many leapfrog steps to explore the posterior.
\begin{table}[t]
\centering
\begin{tabular}{lrr|lrr}
  \hline
Posterior & Steps & Diverg. & Posterior & Steps & Diverg.\\ 
  \hline 
  \small diamonds-diamonds & 970 & 0.00 &
  \small ...covid19...v2 & 300 & 0.20 \\ 
  \small ...covid19...v3 & 260 & 0.22 &
  \small hmm\_gaussian... & 600 & 0.25 \\  
  \small mcycle\_gp... & 660 & 0.13 &
  \small mcycle\_splines... & 1020 & 0.00 \\ 
  \small mnist\_100-nn... & 1020 & 0.00 &
  \small pilots-pilots... & 290 & 0.24 \\ 
  \small ...prophet & 1000 & 0.00 &
  \small soil\_carbon... & 110 & 0.18 \\     
  \small ...RBF...kronecker..gp & 1020 & 0.00 &
  \small uk..state\_space... & 790 & 0.05 \\   
   \hline
\end{tabular}
\vspace{1mm}
\caption{\it Examples of posteriors with difficult geometries for HMC.}
\label{tab_complex_posteriors}
\end{table}

Incorporating complex posteriors without reference draws also includes scenarios where, for example, the current Stan dynamic HMC is inefficient or faces inherent limitations or difficulties. This deliberate inclusion not only prompts and fosters the development of new inference algorithms but also serves as a valuable resource for the community. 

\section{Case study: The Pathfinder algorithm}
\label{sec:case_studies}
The content of \texttt{posteriordb} has already been used in multiple settings to evaluate and develop new algorithms \cite[e.g. see][]{dhaka2020robust}. 
Here, we give an example of how \texttt{posteriordb} was used to develop and evaluate the Pathfinder variational inference algorithm \cite{zhang2022pathfinder}.

The approach utilizes quasi-Newton algorithm---specifically a limited-memory Broyden-Fletcher-Goldfarb-Shanno algorithm (L-BFGS)---to locate low-rank normal approximations to the target distribution along a quasi-Newton optimization path. The covariance for each approximation is provided by the compact and efficient inverse Hessian estimate produced by L-BFGS. Pathfinder evaluates the evidence lower bound (ELBO) in parallel for each normal approximation and returns draws from the approximation which minimizes the ELBO. Pathfinder can quickly find the region of high probability mass from which to draw approximate samples.

Pathfinder was compared to ADVI \cite{kucukelbir2017automatic} and to short MCMC runs, using Stan's implementation of ADVI and HMC/NUTS \cite{stan2021ref}. The latter procedure corresponds to the first stage of MCMC warmup or a variational inference algorithm in its own right, following \cite{hoffman2020black}. The approximation performance was evaluated through the discrete form of the 1-Wasserstein distance and the counts of log density and gradient evaluations.

The Pathfinder algorithm was tested on 20 posteriors from \texttt{posteriordb}. This collection included generalized linear models, hierarchical meta-analysis models, (heteroskedastic) Gaussian processes models, mixture models, differential equation dynamics models, hidden Markov models and time-series models. For each model in \texttt{posteriordb}, the authors ran Pathfinder for 100 runs 
and compared them to the result of 100 runs of 1) Stan phase I adaptation: adaptive Hamiltonian Monte Carlo with Stan's no-U-turn sampler, 2) dense ADVI:  ADVI with a dense covariance matrix, and 3) mean-field ADVI: ADVI with a diagonal covariance matrix.

\begin{figure}[t!]
	\centering{
		\includegraphics[width=\textwidth, keepaspectratio]{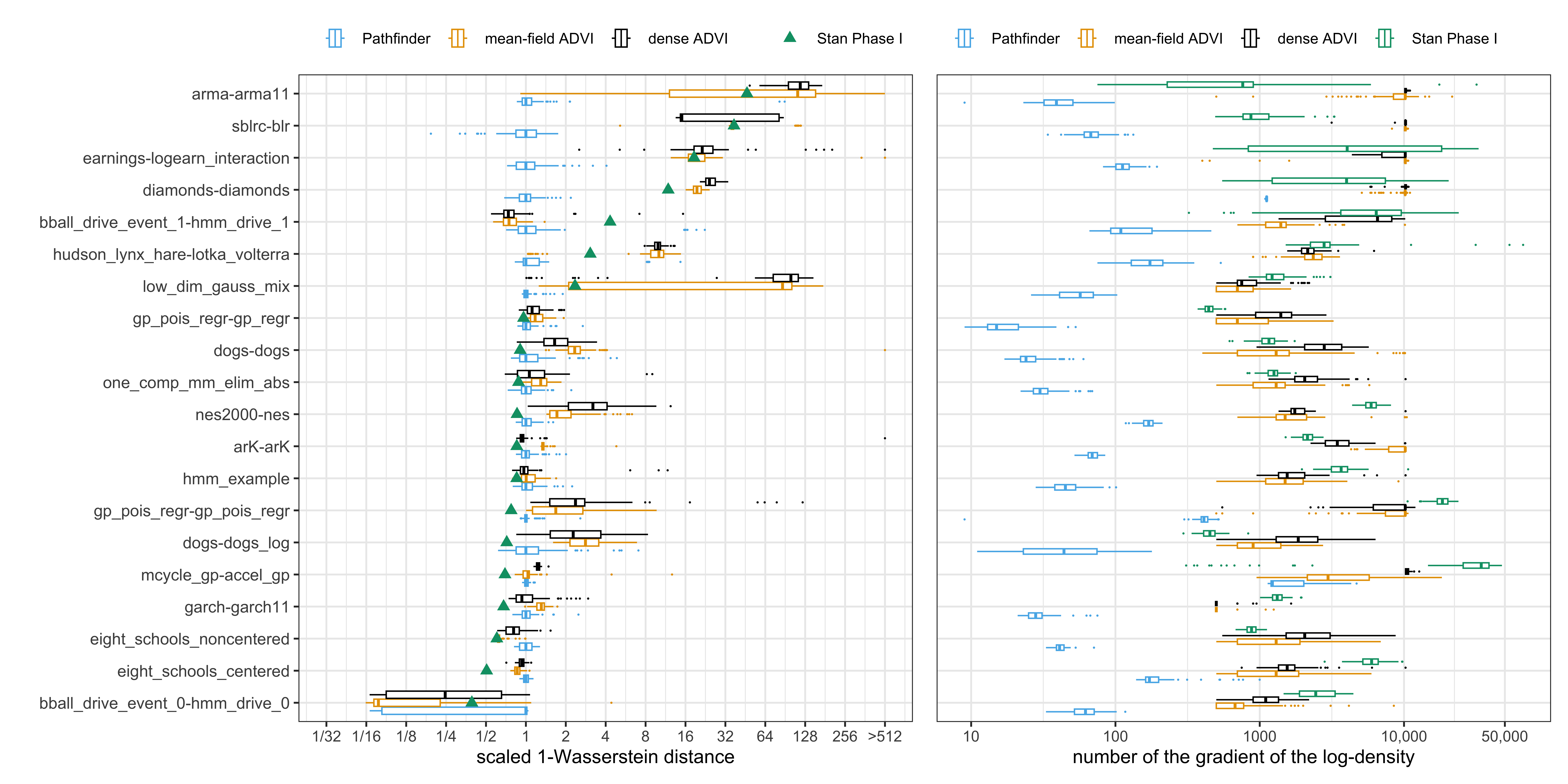}}
	\caption{\it Left panel: Box plots of 1-Wasserstein distances between the reference posterior samples and approximate draws from single-path Pathfinder and ADVI for the 20 models in \texttt{posteriordb}. Each box plot displays 1-Wasserstein distances of 100 independent runs of (single-path) Pathfinder, mean-field ADVI, and dense ADVI. The 1-Wasserstein distance was calculated with 100 approximate draws from the last iteration of 100 runs of Stan's phase I warmup (adaptive HMC). Distances for each model were scaled by the median of the 1-Wasserstein distances for single-path Pathfinder. Right panel: Box plots of the number of gradient evaluations required by the candidate algorithms. Each box plot summarize the cost for 100 independent runs.  
    }\label{fig: pfvsADVI_compar}
\end{figure}

The right panel of Figure~\ref{fig: pfvsADVI_compar} compares the computational efficiency in terms of the number of GE. The implementation-independent computational costs were assessed by the number of LDE and GE. The experiment indicates that Pathfinder requires the lowest computational cost among the evaluated algorithms. In general, the cost of Stan phase I sampler is lower than that of mean-field ADVI, and dense ADVI is the most computationally expensive. Significant differences in computational costs are evident across the various test models and algorithms. 
The left panel of Figure~\ref{fig: pfvsADVI_compar} illustrates a comparison of single-path Pathfinder, ADVI, and Stan's phase I sampler through 1-Wasserstein distances. To adjust for the varying scale of the 1-Wasserstein distances across target densities, the results were rescaled relative to the median of the 100 1-Wasserstein distances for single-path Pathfinder for each model. This allows comparing the ratios of the 1-Wasserstein distances.
Overall, Pathfinder produced relatively lower 1-Wassertein distances compared to ADVI variants. The median 1-Wasserstein distance for mean-field ADVI is more than double that of single-path Pathfinder for 8 (of 20) test models. Dense ADVI showed the greatest variability in the 1-Wasserstein distance to the true posterior. Notably, for the hidden Markov model \verb|bball_drive_event_0-hmm_drive_0|, mean-field ADVI achieves a median 1-Wasserstein distance less than one-tenth that of single-path Pathfinder. This particular model has multiple meaningful posterior modes, and the noise inherent in the stochastic gradient descent approach used by ADVI allows it to escape minor modes that can trap the L-BFGS optimizer used by Pathfinder. Compared to Stan phase I adaptation, single-path Pathfinder exhibited relatively stable performance across challenging posteriors in terms of 1-Wasserstein distances in this experiment. In 7 of the 20 test models, Stan's phase I warmup resulted in 1-Wasserstein distances more than double the median distance of single-path Pathfinder. With the exception of the \verb|bball_drive_event_0-hmm_drive_0| model, the 1-Wasserstein distances for single-path Pathfinder are at most double those of Stan's phase I warmup. 

\section{Discussion and conclusions}
\label{sec:conclusion}
We present \texttt{posteriordb}, a collection of models, data, posteriors, and reference draws for developing PPLs and posterior approximation algorithms. 
In the work with \texttt{posteriordb}, we have gained experience in what is important in benchmarking projects. First, we added many relatively simple posteriors that can be estimated easily using standard default dynamic HMC. In hindsight, more difficult posteriors are more relevant, especially for developing algorithms. Second, labels and information on the posteriors are more important than we first thought. Some posteriors are too complex (multimodal with weak identifiability), leading to very slow computations, while others are too easy. When we used the \texttt{posteriordb} for benchmarking and algorithm development, we realized we needed to pick appropriate posteriors for the experimental goal (e.g., posteriors that are not log-concave may be excluded or be the target of interest). Third, an important conclusion is to separate the model, data, and posteriors to facilitate a broader use and reuse of the components.

\subsection{Limitations and future work}

Presently, the absolute majority of models are Stan models (with some PyMC 
contributions); hence \texttt{posteriordb} currently relies heavily on Stan. However, \texttt{bridgestan} \cite{roualdes2023bridgestan} can access these models' log gradients and densities in many different languages, simplifying the development of algorithms in these languages.
In addition, some posteriors are so challenging that a reference posterior is lacking, making these posteriors difficult to use for benchmarking at the moment. 

We plan multiple developments to \texttt{posteriordb}. First, we intend to add a wider range of posteriors and specifically target more challenging posteriors. Second, we want to incorporate posterior model code from additional PPLs to simplify comparisons and benchmarking also between PPLs. Third, the database should be augmented with predictive distributions or functionality to compute predictive distributions. This would simplify comparisons and diagnostics based on predictive distributions and the development of model assessment diagnostics. Fourth, we have realized the need to identify geometries empirically from draws, for example, funnel-shaped posteriors, posteriors with non-positive-definite Hessians within the set of draws, or posteriors with multiple modes. This is non-trivial today but needs to be simplified further for posterior approximation development.

\subsection*{Acknowledgements}
The research was funded by the Swedish Research Council through grant 2022-03381.

The computations were enabled by resources provided by the National Academic Infrastructure for Supercomputing in Sweden (NAISS), partially funded by the Swedish Research Council through grant agreement no. 2022-06725. 

\bibliographystyle{plain}




\end{document}